\documentclass{PoS}

\usepackage{bm}
\newcommand{\eq}[1]{Eq.(\ref{#1})}
\newcommand{\ud}{\mathrm{d}}

\title{Nuclear Effects in the Deuteron and Global PDF Fits}

\ShortTitle{Nuclear Effects in the Deuteron}

\author{S.I. Alekhin\\
        Institute of High Energy Physics, 142281 Protvino, Moscow Region, Russia\\
        E-mail: \email{sergey.alekhin@ihep.ru}}

\author{S.A. Kulagin\\
        Institute for Nuclear Research of the Russian Academy of Sciences, Moscow 117312, Russia\\
        E-mail: \email{kulagin@ms2.inr.ac.ru}}

\author{\speaker{R. Petti}\\
        Department of Physics and Astronomy, University of South Carolina, Columbia SC 29208, USA\\
        E-mail: \email{roberto.petti@cern.ch}}

\abstract{%
We present a detailed study of nuclear corrections in the deuteron (D) from an analysis of
data from charged-lepton deep-inelastic scattering (DIS) off proton and D,
as well as from dimuon pair production in pp and pD collisions and
$W^\pm$ and the $Z$ boson production at pp (p$\rm \bar p$) colliders.
In particular, we discuss the determination of the off-shell function describing the modification
of parton distributions (PDF) in bound nucleons in the context of global PDF fits.
Our results are consistent with the ones obtained earlier from the study of
the ratios of DIS structure functions $F_2^A/F_2^D$ in nuclei with $A\geq4$,
confirming the universality of the off-shell function.
We also discuss the sensitivity to various models of the deuteron wave function 
and the impact of nuclear corrections on the determination of the $d$ quark distribution.
}

\FullConference{XXIV International Workshop on Deep-Inelastic Scattering\\
		11-15 April 2016\\
		DESY Hamburg, Germany}

\begin{document}

\section{Introduction}

The PDF content for both the proton and the neutron is usually determined from global fits
to experimental data at large momentum transfer $Q^2$.
Traditionally, most of the separation between $d$ and $u$ quark
distributions is obtained by comparing charged-lepton DIS data off proton and deuterium,
the latter being considered as an "effective" neutron target.
The studies available in literature
indicate that nuclear corrections in the DIS off the deuteron
are non negligible and rise rapidly in the region of large Bjorken $x$.
A recent direct measurement of nuclear effects in the deuteron~\cite{Griffioen:2015hxa}
confirms the presence of a few-percent negative correction at $x\sim 0.5-0.6$
with a steep rise at large $x$.
Therefore, if neglected or treated incorrectly, nuclear effects in deuterium can potentially
introduce significant systematic uncertainties or biases in the extraction of the neutron structure functions,
as well as the $d$ quark distribution~\cite{Accardi:2011fa} in the region of large Bjorken $x$.

The deuteron is a weakly bound state of two nucleons with peculiar attributes.
Its dynamics is better understood than the dynamics
of many-particle nuclei, making it a good benchmark for the study of different nuclear effects.
However, since it is considerably different with respect to heavier nuclei, 
we cannot rely on simple extrapolations
from heavier targets based upon individual nuclear properties like nuclear density, atomic weight, etc.
The model of Ref.~\cite{KP04} offers a unified treatment of the deuteron and heavier 
nuclei on the basis of common physics mechanisms~\cite{KP04,KP07,KP10,KP14,Ru:2016wfx}.
In this contribution we report the results of a study of the nuclear corrections in the deuteron
from a combined analysis of various proton and deuteron hard-scattering data.

\section{Nuclear Model and Analysis Framework}
\label{sec:model}

The nuclear structure functions for the inelastic scattering off the deuteron
involve a number of different contributions
(for brevity, we suppress explicit dependencies on $x$ and $Q^2$)\cite{KP04}:
\begin{eqnarray}
\label{F2D}
F_2^D &=& \left\langle F_2^{N} \right\rangle + \delta_\mathrm{MEC} F_2^D + \delta_\mathrm{coh} F_2^D,
\\
\left\langle F_2^{N} \right\rangle &=&
		\int\frac{\ud^3\bm p}{(2\pi)^3} \left|\Psi_D(\bf p)\right|^2 \left(1+\frac{p_z}{M}\right)
			F_2^N(x',Q^2,p^2)
\label{F2D-incoh}
\end{eqnarray}
%
where \eq{F2D-incoh} implies the convolution of the 
square of the deuteron wave function 
$\Psi_D(\bf p)$, describing the nucleon momentum distribution in the deuteron, with the 
structure function $F_2^N(x',Q^2,p^2)$ of the bound nucleon with 
four-momentum $p=(M_D-\sqrt{M^2+\bm p^2},\bm p)$ with $M_D$ and $M=(m_p+m_n)/2$
the deuteron and the nucleon mass, respectively.
\footnote{In \eq{F2D-incoh} we drop power terms $\sim Q^{-2}$ for illustration purpose only.
For the full expression see Ref.\cite{KP04}}
The terms $\delta_\mathrm{MEC} F_2^D$ and $\delta_{\rm coh} F_2^D$ are the corrections
arising from nuclear meson exchange currents (MEC) and
coherent interactions of the intermediate virtual boson with nuclear target.

The first term in Eq.(\ref{F2D}) represents the contribution from the incoherent scattering 
off the bound nucleon and dominates at $x>0.1$.
In the following we will consider several deuteron wave functions $\Psi_D(\bm p)$
corresponding to different models for the nucleon-nucleon potential:
Paris~\cite{Lacombe:1980dr}, Bonn~\cite{Machleidt:1987hj,Machleidt:2000ge}, AV18~\cite{Veerasamy:2011ak},
WJC-1 and WJC-2~\cite{Gross:2008ps,Gross:2010qm}.
These wave functions are constrained by high-precision fits to
nucleon-nucleon scattering data at low energies.
The nucleon structure function includes a leading twist (LT) term corrected for the target mass
effect (TMC) \cite{Georgi:1976ve} and also a phenomenological higher-twist (HT) power correction, $F_2^N(x,Q^2)=F_2^\mathrm{LT,TMC}(x,Q^2)+H_2(x)/Q^{2}$.

The structure function of the bound nucleon $F_2^N$ entering Eq.(\ref{F2D}) explicitly
depends on the nucleon invariant mass squared $p^2$.
The $p^2$ dependence of the structure function has two different sources \cite{KP04}:
(i) the kinematic target mass correction, which generates terms of the order of $p^2/Q^2$;
and (ii) the dynamical off-shell (OS) dependence of the LT structure functions.
We evaluate the off-shell dependence of the target-mass correction terms of 
Ref.\cite{Georgi:1976ve} by replacing $M^2\to p^2$.
In order to address the off-shell dependence of the LT structure function,
we consider the latter in the vicinity of the mass shell $p^2\approx M^2$:
\begin{equation}
\label{eq:sf-os}
F_2^\mathrm{LT}(x,Q^2,p^2)=F_2^\mathrm{LT}(x,Q^2)\left(1+\delta f(x,Q^2) v\right),
\end{equation}
where we keep the leading term in the series expansion with respect to the virtuality
$v=(p^2-M^2)/M^2$. The function
$\delta f = \partial\ln F_2^\mathrm{LT}/\partial\ln p^2$ describes the
modification of the nucleon structure functions and PDFs in the vicinity of the mass shell.
This function does not contribute to the cross section of the
physical nucleon, but it is relevant only for the bound nucleon and describes its
response to the interaction in a nucleus.

The function $\delta f$ was determined phenomenologically
from a global analysis of data on the ratio of DIS structure functions $F_2^A/F_2^B$~\cite{KP04} 
between two different nuclear targets.
The predictions of Ref.\cite{KP04,KP10} describe the observed $x$, $Q^2$, and $A$ dependencies
of all available DIS data with high accuracy  for a wide range of
target nuclei from ${}^3$He to ${}^{208}$Pb.
In addition, the predictions of the same model are in a very good agreement
with data on the Drell-Yan production of the dimuon pair off various nuclear targets~\cite{KP14},
as well as the differential cross-sections
and asymmetries of $W^\pm$ and $Z$ boson production in pPb collisions at the LHC~\cite{Ru:2016wfx}.

In this analysis we use the NNLO approximation in the QCD perturbation theory
to calculate the parton cross-sections entering the LT terms for
the hard interaction processes considered.
The parametrization of PDFs follows Ref.\cite{Alekhin:2015cza}
at the starting scale $\mu^2=Q^2_0=9$ GeV$^2$.
We also include two twist-4 terms for the isoscalar nucleon structure functions
with the coefficients $H_2^N$ and $H_T^N$
parameterized as model-independent cubic spline functions of the Bjorken variable $x$.
In this global PDF analysis we do not consider
the meson exchange currents and coherent nuclear effects
(shadowing) for the deuteron, since their impact
is negligible in the kinematic coverage of our analysis.
The only free parameters entering the nuclear corrections are the ones describing the
off-shell function $\delta f(x)$, which are extracted simultaneously with the PDFs and
the HT terms in our global fits to the data. To this end, we use a model-independent
parameterization with generic second and third order polynomials for $\delta f(x)$.
The data samples used in the analysis include the cross sections of inclusive DIS off proton and deuterium
measured by the SLAC E49, E87, E89, E139, E140 experiments, the CERN BCDMS and NMC experiments,
the HERA H1 and ZEUS experiments,
the cross sections of the
Drell-Yan process in pD and pp collisions from the Fermilab E866 experiment,
charm production data in neutrino DIS from the NOMAD and CHORUS experiments,
the $W^\pm$ and $Z$ boson production cross sections and the asymmetries measured
at the Tevatron and the LHC by D0, ATLAS, CMS and LHCb experiments
(for more detail about this analysis see \cite{AKP-to-appear}).

\section{Results and Discussion}
\label{res}

\begin{figure}[htb]
\centering
\includegraphics[width=0.8\textwidth,height=0.33\textheight]{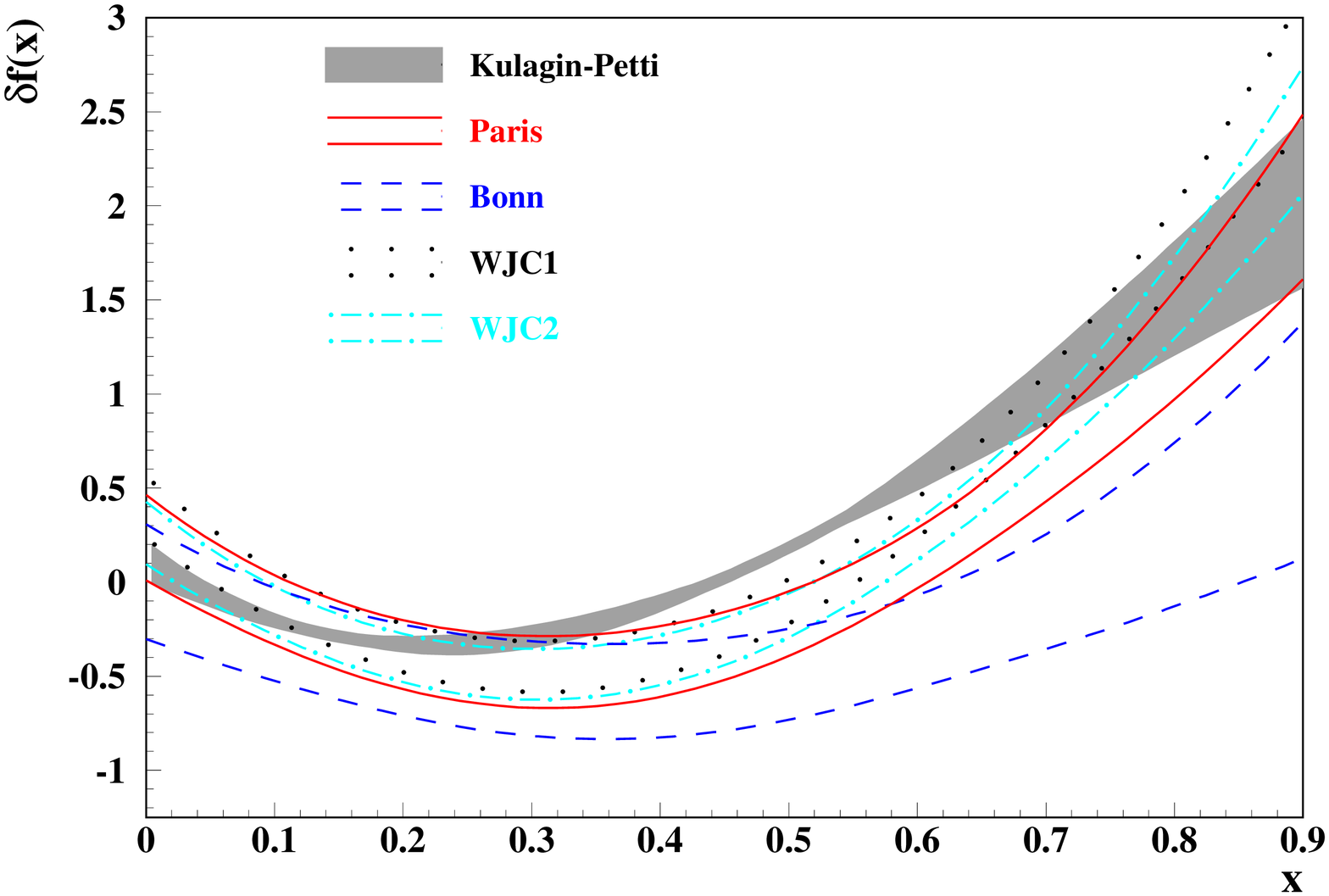}
\caption{%
Comparison of the off-shell functions $\delta f(x)$ ($\pm 1 \sigma$ uncertainty bands)
extracted within our global PDF fit by using different models for the deuteron wave functions as input.
The corresponding uncertainty band obtained from heavy target data ($A \geq 4$) in Ref.~\cite{KP04} is
also shown.
\label{fig:dfwf}}
\end{figure}

The results on the function $\delta f(x)$ from the present analysis are shown
in Fig.\ref{fig:dfwf}. We note that a simultaneous
extraction of the proton PDFs, HT terms and the off-shell function $\delta f$
is possible because of the different
$Q^2$ dependence of these three contributions and a wide $Q^2$ coverage of the data.
In general, the nuclear corrections in the deuteron are
partially correlated with the $d$-quark distribution. In order to reduce this correlation, the data
on the Drell-Yan reaction and the $W^\pm$ boson production from $pp$ and $p\bar{p}$ colliders is crucial.
In particular, the use of recent combined D0 data and the LHC data from CMS and LHCb,
reaching the values of $x\sim 0.8$ due to a wide rapidity coverage,
allows the determination of the proton PDFs with a precision comparable to that
reached in traditional PDF analyses with data from fixed-target DIS experiments.

Figure~\ref{fig:dfwf} indicates that the extraction of $\delta f$ at large $x$
is sensitive to the choice of the model of the deuteron wave function.
This effect is directly related to the spread in the strength of the high momentum 
component among the various deuteron wave functions.
The contribution from the term $F_2^N\delta f v$ is indeed quite sensitive to
the high momentum tail of the wave function, which affects both the $x$ and $Q^2$
dependence.
The general trend observed from Fig.\ref{fig:dfwf} indicates that harder
deuteron momentum distributions correspond to higher values of the resulting $\delta f$ at large $x$.
We also note that, while our results obtained with the Paris, AV18, WJC-1, and
WJC-2 wave functions are consistent within the corresponding uncertainties,
the use of the softer Bonn wave function produces a somewhat larger deviation.

As summarized in Sec.\ref{sec:model},
a more precise determination of the off-shell function $\delta f(x)$ was
obtained in Ref.\cite{KP04} from an analysis of the ratio of nuclear DIS structure functions 
$F_2^A/F_2^B$. In order to further study
the sensitivity to the nuclear smearing, we repeat the standalone extraction of
$\delta f(x)$ following Ref.\cite{KP04} after rescaling the high-momentum part of the
nuclear spectral function
by the ratio of the various deuteron wave functions used for this study.
The results are shown in Fig.\ref{fig:dfwf}.
The gray area represents the $\pm 1 \sigma$ uncertainty band,
including a number of sources of theoretical (systematic) uncertainties 
related to the variation of the nuclear spectral function, the deuteron wave function,
the choice of the functional form to parametrize the function $\delta f$,
the proton PDF uncertainties, and other nuclear corrections like MEC and NS.
We observe a dramatic reduction of the uncertainties
related to the choice of the nuclear spectral function and/or of the deuteron wave function
with respect to the present analysis. This reduction can be explained by the different observables
considered in the two independent extractions of $\delta f$.
In the global PDF fit we use the absolute DIS
cross-sections off the deuteron, while in the standalone determination of Ref.\cite{KP04}
we consider only \emph{ratios} of nuclear structure functions.
Many model uncertainties largely cancel out in such ratios.
Similarly, the data sets used in Ref.\cite{KP04}
are characterized by a higher precision with respect the to the deuteron data used in a global PDF analysis.

The function $\delta f$ extracted from a global PDF fit is consistent with the more precise result
obtained from the analysis of the ratios $F_2^A/F_2^B$ described in Ref.\cite{KP04}.
Since we are using a generic polynomial to parameterize
$\delta f$, no functional form bias is present in this comparison.
The agreement between the two independent determinations provides additional evidence in favor of
the universality of the off-shell function $\delta f$.
This result validates the unified treatment of the deuteron and
heavier nuclei provided by the model of Ref~\cite{KP04}.

\begin{figure}[htb]
\centering
\includegraphics[width=0.8\textwidth,height=0.33\textheight]{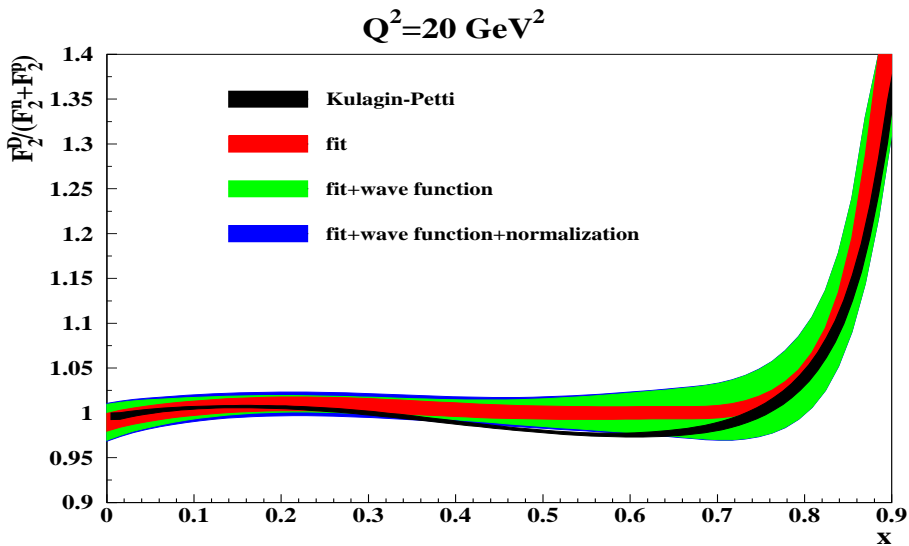}
\caption{%
Summary of the ratio $F_2^D/F_2^N$ obtained from this analysis with the corresponding
$\pm 1 \sigma$ bands including the uncertainty from the fit as well as the systematic
uncertainties related to the choice of the deuteron wave function (central green band) and to the
normalization of the BCDMS deuteron data (blue outer band).
The $\pm 1 \sigma$ uncertainty band obtained from heavy target data ($A \geq 4$) in Ref.\cite{KP04} is
also shown as a solid black area.
\label{fig:f2d-final}}
\end{figure}

The differences related to the choice of the deuteron wave function
in the global PDF fit appear to be substantially reduced in the ratio $F_2^D/F_2^N$
with respect to the off-shell functions $\delta f$ shown in Fig.\ref{fig:dfwf}.
We quantify this uncertainty by comparing the results obtained using the Paris,
Bonn, AV18, WJC-1 and WJC-2 wave functions  and by
considering the largest differences observed.
We also found that the normalization of the deuteron BCDMS data introduces an
additional uncertrainty in the extraction of $\delta f$
and on the $F_2^D/F_2^N$ ratio.
Figure~\ref{fig:f2d-final} summarizes our final results
for the ratio $F_2^D/F_2^N$ as well as the impact of the uncertainties from our global fit,
from the choice of deuteron wave function, and from the normalization of the BCDMS deuteron data.

%

We can exploit the higher precision offered by the DIS data off heavier nuclear
targets ($A\geq4$) to fix the function $\delta f$ used in
global PDF fits following Ref.\cite{KP04}.
Figure~\ref{fig:f2d-final} illustrates the corresponding reduction of the overall uncertainties
on the deuteron nuclear corrections. This result allows, in turn, a more precise determination 
of the $d$ quark distribution at large $x$ values from DIS data off the deuteron.  
Within a simple single-scale model, in which the quark momentum distributions in the
nucleon are functions of the nucleon radius~\cite{KP04}, the universal off-shell function $\delta f$
determines an increase of the nucleon core radius of about 2\% in the deuteron, assuming
an average virtuality of $\bar v= -0.045$.

Our results on the off-shell correction differ from the ones reported in
Refs.\cite{Accardi:2011fa,Accardi:2016qay} on the basis of a similar formalism. The analysis of
\cite{Accardi:2011fa} uses a modified model proposed in \cite{KP04} to relate
the off-shell function $\delta f$ to a change in the nucleon confinement radius
in the nuclear medium. The analysis \cite{Accardi:2016qay} follows more closely the
model \cite{KP04} and determines the off-shell function $\delta f$ from a global
PDF fit to the deuteron and the proton data. The differences in the results with respect to our
study can be attributed to the model implementation and to the details of the fit procedure.
In particular, it is worth noting that the off-shell effect on the target-mass 
correction~\cite{Georgi:1976ve}
was not taken into account~\cite{AccardiDIS16} in~\cite{Accardi:2011fa,Accardi:2016qay}.
This effect results in significant corrections in the large $x$ region. Additional differences are
present in the treatment of the HT terms and in the data sets used.


We thank F. Gross and W. Polyzou for providing the parameterizations of the WJC and AV18 deuteron
wave functions, respectively.
We thank W. Melnitchouk and A. Accardi for discussions.
The work of S.K. was supported by the Russian Science Foundation grant No.~14-22-00161.
R.P. was supported by the grant DE-FG02-13ER41922 from the Department of Energy, USA.

\end{document}